\documentclass[twocolumn]{aastex61}
\usepackage{amssymb}
\usepackage{graphicx,epstopdf}
\usepackage{natbib}

\begin{document}

\newcommand{\wn}{cm$^{-1}$}
\newcommand{\amod}{NH$_3$D$^+$}
\newcommand{\amodd}{NH$_2$D$_2^+$}
\newcommand{\amoddd}{NHD$_3^+$}
\newcommand{\amoh}{NH$_4^+$}

\shorttitle{Accurate rotational  rest frequencies for ammonium ion isotopologues}
\shortauthors{Dom\'enech, Schlemmer and Asvany}
\submitjournal{The Astrophysical Journal}
\accepted{September 4, 2018}

\title{Accurate rotational  rest frequencies for ammonium ion isotopologues }

\author[0000-0001-8629-2566]{Jos\'e L. Dom\'enech}
\affiliation{Instituto de Estructura de la Materia (IEM-CSIC), Serrano 123, E28006 Madrid, Spain}
\author[0000-0002-1421-7281]{Stephan Schlemmer}
\affiliation{I. Physikalisches Institut, Universit\"at zu K\"oln, Z\"ulpicher Str.~77,
50937 K\"oln, Germany}
\author[0000-0003-2995-0803]{Oskar Asvany}
\affiliation{I. Physikalisches Institut, Universit\"at zu K\"oln, Z\"ulpicher Str.~77,
50937 K\"oln, Germany}
\correspondingauthor{Oskar Asvany}
\email{asvany@ph1.uni-koeln.de}

\begin{abstract}
We report  rest frequencies for rotational transitions of the deuterated  ammonium isotopologues \amod, \amodd\ and \amoddd,
measured in a cryogenic ion trap machine.
For the symmetric tops \amod\ and \amoddd\ one and three transitions are detected, respectively,  
and five transitions are detected for the asymmetric top \amodd.
While the lowest frequency transition of \amod\  was already known in the laboratory and space, 
this work enables the future radio astronomical detection of the two other isotopologues.
\end{abstract}


\section{Introduction}
Nitrogen is one of the most abundant elements in the local universe, and has a notably rich chemistry, with more than seventy nitrogen-containing 
molecules identified in space to date \citep{CDMSWeb2}.  
Two of the most abundant 
nitrogen-bearing molecules in the interstellar medium (ISM) are N$_2$ and NH$_3$, which are predicted to be present in many 
different media, from cold dark clouds (see, e.g., \citet{Nejad1990}) or protostellar cores (see, e.g., \citet{Aikawa2008}) 
to active galactic nuclei (see, e.g., \citet{Harada2010}). 
N$_2$, being a homonuclear diatomic molecule, can not be observed through its (electric-dipole) rotation or vibration transitions.
It can be observed, however, through electronic transitions in the far-ultraviolet, with only  one direct observation reported by \cite{Knauth2004},
or by its frequently used proxy, the  diazenylium ion (N$_2$H$^+$).
Ammonia, in turn, is the first polyatomic molecule identified in space \citep{Cheung1968a}, 
and since then has been observed in many environments. 
To date, all deuterated isotopologues of ammonia, NH$_2$D, NHD$_2$ and ND$_3$ have been observed (see e.g. \citet{Harju2017} 
for a recent account), as well as $^{15}$NH$_3$ and $^{15}$NH$_2$D \citep{Gerin2009}.

The astrophysical relevance of the ammonium ion, \amoh, stems from its role as gas phase forebear of the ammonia molecule 
through its dissociative recombination.  The full family of nitrogen hydrides is initiated by the reaction N$^+$+H$_2\rightarrow$ NH$^+$ + H \citep{zym13}.  
Subsequent exothermic H$_2$ abstraction reactions lead to the formation of NH$_2^+$, NH$_3^+$ and NH$_4^+$, and their recombination 
with electrons form the neutral hydrides NH, NH$_2$ and NH$_3$.  Ammonia is supposed to be mainly depleted onto grains at the 
temperatures of cold dark clouds ($T\sim10$~K).  On the other hand, the grains can free NH$_3$ molecules upon heating by 
shocks or in irradiated regions, increasing its gas-phase concentration, and leading to appreciable quantities of \amoh, 
formed by the proton transfer from H$_3^+$ to NH$_3$.  Actually, NH$_3$ has one of the highest proton affinities (PA) of 
simple interstellar molecules (PA[NH$_3$]=8.85 eV) including H$_2$ (which has PA[H$_2$]=4.39 eV), so \amoh, once formed, 
remains stable against further collisions with H$_2$.  

\amoh\ has tetrahedral symmetry with no permanent electric dipole moment. Therefore, it is untraceable by 
radio astronomy through its rotational transitions.  On the other hand, the deuterated isotopologues \amod, \amodd\ 
and \amoddd\  have sizeable permanent electric dipole moments (approximately 0.26 D, 0.29 D and 0.24 D, respectively) 
due to the separation between the center of charge (which remains centered on the central N atom) and the center of 
mass (which is displaced towards the D atoms),
thus making their detection feasible.
Indeed, the detection of ammonium in space was claimed through the assignment of an emission line centered 
at 262817~GHz (observed both in Orion IRc2 and in Barnard B1-bS) to the $1_0-0_0$ transition of \amod\ \citep{Cernicharo2013}.  
The laboratory rest frequency was derived from the analysis of the high resolution infrared spectrum of the $\nu_4$ band, 
originally made by \citet{Nakanaga1986} and considerably improved by \citet{dom13}.  The rest frequency was later confirmed, 
and its uncertainty substantially decreased, by the direct measurement of the rotational transition by \citet{sto16}.

Besides their importance as tracers of \amoh, the deuterated isotopologues are also relevant regarding the subject of deuterium fractionation.  
In cold dark clouds, when CO freezes onto grains and its abundance falls below that of HD, deuteration of other species becomes efficient.  
It can be explained by the exothermicity  ($\sim230$~K) of the reaction ${\rm H_3^+ + HD \rightarrow H_2D^+ + H_2}$ and subsequent 
deuteron transfer to other molecules.  Similarly, the exothermic reactions of H$_2$D$^+$ and HD$_2^+$ with HD efficiently produce HD$_2^+$ and D$_3^+$
\citep{Hugo2009} in case of extreme depletion, and subsequent deuteron transfer to ammonia, e.g.\
${\rm NH_3 + H_2D^+ \rightarrow NH_3D^+ + H_2}$, lead to the deuterated forms of \amoh.
Furthermore, the zero point energy differences of other 
reactions involving H or D atoms at temperatures of cold dark clouds ($T\sim10$~K) also enhance deuteration \citep{Millar2003}.  
For the case of nitrogen, the reaction ${\rm N^+ + HD \rightarrow ND^+ + H}$
has a lower endothermicity than the corresponding reaction with H$_2$ mentioned above \citep{gro16}, 
leading to an enrichment in deuterated ammonium and ammonia isotopologues.  
The detected abundances of deuterated variants of ammonia are orders of magnitude higher compared to what 
is expected based on the [D]/[H] abundance ratio (2.35$\times10^{-5}$ in the local universe \citep{Linsky2006}).  
Detection of other deuterated ammonium isotopologues besides \amod\ will clearly help to constrain and understand 
the conditions of formation and distribution of ammonia molecules, and, possibly, other nitrogen-containing prebiotic species.

The only previous spectroscopic laboratory data on the doubly and triply deuterated species (\amodd\ and \amoddd) 
that we are aware of are the recent publication of the analysis of the $\nu_1$ and $\nu_6$ infrared 
bands of \amodd\ \citep{Chang2018a}, and the communication of preliminary results on the analysis of 
the $\nu_1$ band of \amoddd\  \citep{Chang2013b}, 
both recorded at high resolution in a supersonic slit-jet discharge with an infrared  difference-frequency laser spectrometer.
In this work, we present direct and accurate laboratory measurements of the lowest frequency rotational 
transitions of \amod, \amodd\ and \amoddd.  The fundamental rotational frequency for \amod\ has already been published previously, 
and we only confirm its value, while our reported values for the other two isotopologues represent the first 
available sub-mm wave data.  

\section{Experimental methods}

The rotational transitions of the deuterated ammonium isotopologues have been measured in the K\"oln 
laboratories  exploiting the rotational state dependence of 
the attachment of He atoms to cations (\cite{bru14,bru17,jus17,dom17,dom18}). 
The experiment was performed in the 4~K trapping machine COLTRAP described by \cite{asv10,asv14}.
The  ions were generated in a storage ion source by 
bombarding the precursor gas mixture with  electrons (energy 30-40~eV).  
For generating \amod\ and \amodd,  NH$_3$ (Messer, 99.8~\% purity) and a 1:5 mixture of  D$_2$ (Linde, 99.8~\%) and He (Linde 99.999~\%)
were admitted to the ion source via two separate leakage valves. 
The approximate   proportions were 1:1:5 for \amod\ and   1:2:10  for \amodd.
For \amoddd\ we applied a similar mixture using  ND$_3$ (Campro Scientific, 99~\% atom D) and H$_2$ (Linde, 99.9999\%). 
A pulse of several ten thousand mass-selected parent ions was injected into 
the 4~K cold 22-pole ion trap  filled with about 10$^{14}$ cm$^{-3}$ He.
During the trapping time of  750~ms,  cation-helium complexes formed 
by three-body collisions. The detection of the resonant absorption of the admitted 
cw (sub)millimeter radiation by the trapped cold {\em parent} cations  
was achieved by observing the decrease of the number of cation-helium complexes formed.   
For example, for recording a rotational transition of \amodd, the \amodd$-$He complexes are counted (typical counts are on the order of 2000) as a function
of the (sub)millimeter-wave frequency.
The (sub)millimeter-wave radiation was supplied by a rubidium atomic clock-referenced synthesizer (Rohde\&Schwarz SMF100A) driving a  multiplier chain source
(Virginia Diodes, Inc.), covering the range 80-1100~GHz.


\section{Results}

\begin{figure}[h!]
\begin{center}
\vspace{1cm}
\includegraphics[width=0.75\columnwidth]{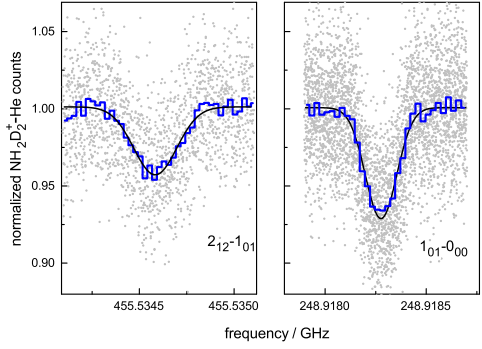}
\caption{\label{figure:lines} Example measurements of the $1_{01}-0_{00}$ and $2_{12}-1_{01}$ rotational transitions of \amodd, 
recorded as depletion signal of the normalized \amodd$-$He counts.   
Grey dots are single measurements, the blue line is the average binned in 25 kHz steps, and the black trace is a Gaussian fit.
}
\end{center}
\end{figure}

We started our measurements by optimizing the experimental conditions using the known $1_{0}-0_{0}$ line of \amod, 
which has been measured by \cite{sto16} with the same ion trap technique. We confirm their value for $1_{0}-0_{0}$, as well as 
the inability of the applied setups to detect the $2_{0}-1_{0}$ and  $2_{1}-1_{1}$ transitions. All experimental values of this work are 
summarized in Table~\ref{tab1}.\\
%
%
For \amodd, we based our search on the predictions from the 
high resolution infrared vibration-rotation spectra of the $\nu_1$ and $\nu_6$ bands 
recently published by \citet{Chang2018a}. In total, we searched for eight rotational lines of \amodd\ of which five were detected. 
All measured and predicted transitions  are included in Table~\ref{tab1}.
Figure~\ref{figure:lines} shows two example measurements for \amodd.
All lines have been measured at least 7 times and fitted to Gaussian functions, from which line centers and widths were determined. The values 
quoted in Table~\ref{tab1} represent the combined mean and standard deviation of all measurements. 
During the  measurements, care was taken to avoid power broadening. By this, the linewidths are   
determined only by the Doppler broadening due to the kinetic temperature of the ions in the trap (nominal temperature $T=4$~K), and also by a small 
contribution from the unresolved hyperfine splitting due to the quadrupole moments of the $^{14}$N and $^2$H nuclei, both with spin I=1.
Effectively, we measure linewidths corresponding to $T= 10 - 20$~K for \amodd, depending on the observed line. The  $1_{0}-0_{0}$ line of 
\amod\ seems to be more affected by hyperfine splitting, with an effective temperature $T \approx 30$~K. \\

For \amoddd, our search was guided by mass-scaling the rotational constants from the other isotopologues, 
as well as 
{ by an estimation based on the available IR-derived rotational constants}
from the Nesbitt group (\cite{Chang2013b}). In contrast to \amod, we were able to measure not only the  fundamental $1_{0}-0_{0}$ for \amoddd, 
but also detected the $2_{0}-1_{0}$ and  $2_{1}-1_{1}$ transitions. The reason is a more advantageous partition function for \amoddd,
as well as more available millimeter-wave power at the lower transition frequencies of the heavier \amoddd.

\begin{table}[h]  
\caption{\label{tab1} Frequencies of pure rotational transitions  (in MHz) of  deuterated ammonium isotopologues.  
The final error is given in parentheses in units of the last digit. }
\begin{center}
\begin{tabular}{lrclr@{}lr@{}l}
\hline
\amod   & $J'_K$&$\leftarrow$&$J''_K$                  &  \multicolumn{2}{c}{this work}         &  \multicolumn{2}{c}{former work$^a$}         \\
        & $1_0$&$\leftarrow$&$0_0$                     &             262816&.8864(8)           &     262816&.904(15)                          \\
\hline
\amodd  & $J'_{ K_a' K_c'}$&$\leftarrow$&$J''_{K_a'' K_c''}$ &  \multicolumn{2}{c}{this work}        &   \multicolumn{2}{c}{former prediction$^b$}         \\
        & $1_{10}$&$\leftarrow$&$1_{01}$                 &              42273&.4715(47)$^c$         &       42270&(6)         \\   
        & $1_{11}$&$\leftarrow$&$0_{00}$                 &             248918&.2760(8)          &       248924&(6) \\
        & $2_{02}$&$\leftarrow$&$1_{11}$                 &             412755&.5374(27)          &            &        \\
        & $2_{12}$&$\leftarrow$&$1_{01}$                 &             455534&.5852(16)          &      455540&(12)\\
        & $2_{21}$&$\leftarrow$&$1_{10}$                 &             540046&.0112(75)          &            &    \\
        & $2_{20}$&$\leftarrow$&$1_{11}$                 &             560809&.9582(37)          &            &   \\
        & $3_{03}$&$\leftarrow$&$2_{12}$                 &             632874&.2(17)$^c$         &            &   \\
        & $3_{13}$&$\leftarrow$&$2_{02}$                 &             656593&.10(82)$^c$         &            &        \\
\hline
\amoddd & $J'_K$&$\leftarrow$&$J''_K$                  &  \multicolumn{2}{c}{this work}        &             &          \\
        & $1_0$&$\leftarrow$&$0_0$                     &              222228&.9432(7)         &              &          \\
        & $2_0$&$\leftarrow$&$1_0$                     &              444415&.7929(42)        &             &          \\
        & $2_1$&$\leftarrow$&$1_1$                     &              444421&.2980(18)         &             &          \\
\hline
\end{tabular}
\end{center}
$^a$ \cite{sto16}\\
$^b$ \cite{Chang2018a}\\
$^c$ our prediction
\end{table}

%


\section{Spectroscopic parameters}
As accurate spectroscopic parameters and predictions for the ground state of \amod\ were already given by \cite{sto16} and \cite{dom13}, 
we treat here only the isotopologues \amodd\ and \amoddd.   
%
For \amodd, the measured frequencies of the pure rotational lines collected in Table~\ref{tab1} were fit together with the 
ground state combination differences derived from \citet{Chang2018a}(two duplicated entries had to be trimmed) using the program PGOPHER \citep{Wes17}, 
rendering the final set of parameters for the ground state given in Table~\ref{tab2}.   Both sets of 
data were weighted according to their different uncertainties, in the kHz range for the present measurements, 
and $\sim7$~MHz for the IR measurements, resulting in a weighted standard deviation of the fit $\sigma_w=1.02$.  The accuracy of the spectroscopic parameters of this work is marginally improved (a factor of $\sim2-3$) with respect to the former work, since only five rotational lines could be recorded in high resolution but eight parameters are fit.    It may be also noted that the $\delta_k$ parameter is not significantly different from zero in this fit.\\
%
For \amoddd,  we only have 
{ the three presently measured transition frequencies available}
therefore we cannot provide a least squares fit.  
Since, furthermore, the transitions obey $\Delta K=0$, we cannot determine $C$ and $D_K$.
With the three experimental frequencies we have determined $B$, $D_J$ and $D_{JK}$, shown in Table~\ref{tab2} 
together with the values quoted by \citet{Chang2013b}.  Our quoted errors are obtained propagating the experimental uncertainties
{ through the closed linear relations existing between the three
observed frequencies and the three retrieved constants. We have assumed that there is
no correlation between the experimental frequencies, thus the uncertainties are just
added in quadrature with the appropriate coefficients.}
\begin{table}[h]
\caption{\label{tab2} Derived spectroscopic parameters (in MHz) from a fit to all measured transitions.
    Numbers in parentheses are one standard deviation in units of the last digit.}
\begin{center}
\begin{tabular}{llr@{}lr@{}l}
\hline
\amodd & Parameter           & \multicolumn{2}{c}{this work}      & \multicolumn{2}{c}{former work$^a$}  \\
\hline
       &  $A$                &           145600&.69(33)        &       145601&.7(12)               \\    
       &  $B$                &           118963&.31(77)        &       118966&.3(12)           \\ 
       &  $C$                &           103324&.30(63)        &       103328&.6(18)            \\ 
       &  $\Delta_J$         &           1&.698(46)            &            2&.10(9)                        \\    
       &  $\Delta_{JK}$      &          -0&.85(24)             &           -1&.9(4)           \\ 
       &  $\Delta_K$         &           3&.08(24)             &            3&.87(3)           \\ 
       &  $\delta_J$         &           0&.371(15)            &            0&.18(6)         \\ 
       &  $\delta_K$         &           0&.01(20)             &            0&.9(3)           \\ 
\hline
\amoddd & Parameter            & \multicolumn{2}{c}{this work}       & \multicolumn{2}{c}{former work$^b$}  \\
\hline
      &  $B$                &          111117&.9794(6)                        &       111120&.9(8)        \\ 
      &  $D_J$              &               1&.7539(2)                        &            2&.01(7)         \\  
      &  $D_{JK}$           &              -1&.3763(11)                       &           -2&.0(2)        \\   
\hline
\end{tabular}
\end{center}
$^a$ \cite{Chang2018a} \\
$^b$ \cite{Chang2013b}
\end{table}

\section{Conclusion and Outlook}
We have measured, with kHz-level accuracies, rotational transitions originating from the lowest energy levels of the deuterated 
ammonium isotopologues \amod, \amodd\ and \amoddd.  Prior to this work, \amod\ was already identified in Orion IRc2 and 
Barnard B1b \citep{Cernicharo2013}.  For the \amodd\ and \amoddd\ isotopologues these rest frequencies will facilitate the 
search for these ions in the interstellar medium.  Cold dark clouds, highly depleted in C and O are the kind of sources 
where high deuterium fractionation is expected.
The cold protostellar core B1b is a source where the abundance of deuterated isotopologues of ammonia  and the 
diazenylium ion (N$_2$H$^+$) is quite remarkable, as is the presence of other multiply deuterated species, like D$_2$CS \citep{Marcelino2005} 
or D$_2$S \citep{Vastel2003}.  The chemical model used in the interpretation of \amod\ observations in B1b predicted that 
the abundance of \amodd\ could be just a factor of two lower than that of \amod.  Therefore, one could expect the detection of 
more deuterated species of the ammonium ion in this or other cold dark clouds.  We note, however, that the partition 
function of \amodd, being an asymmetric top, is more unfavorable, and, even at the low temperature of B1b ($T\sim12$~K), 
spectral dilution is to be expected.  The slightly higher electric dipole moment (0.29~D vs 0.26~D) just marginally improves this situation.  
Although the necessary sensitivity for this detection seems demanding,  the ever increasing capabilities of receivers and 
observatories will likely make it possible in a not distant future.  \amoddd\ has a smaller abundance and dipole moment, 
but it has higher symmetry (it is a symmetric top, like \amod) and, therefore, a more favorable partition function, 
so it may also be soon within the reach of observational radio astronomy.  
In any case the frequencies provided in this work will enable those detections.

\acknowledgments
This work (including a research stay of J.L.D. in K\"oln) has been  supported by the Deutsche
Forschungsgemeinschaft (DFG) via SFB 956 project B2
and the Ger\"atezentrum "Cologne Center for Terahertz Spectroscopy".
J.L.D. acknowledges partial financial support from the Spanish MINECO through grant
FIS2016-77726-C3-1-P and from the European Research Council through grant
agreement ERC-2013-SyG-610256-NANOCOSMOS.
The authors gratefully acknowledge the work done over the last years 
by the electrical and mechanical workshops of the I. Physi\-kali\-sches Institut.
We thank  Marius Hermanns, Frank Lewen and Matthias T\"opfer for assistance in the measurements.
\software{PGOPHER \citep{Wes17}}


\bibliography{nh2d2+,LIRTRAP,misc}

\end{document}